\documentclass[conference,a4paper]{IEEEtran}

\usepackage{amsmath}
\usepackage{amssymb}
\usepackage{latexsym}
\usepackage{array,arydshln}
\usepackage{multirow}
\usepackage{graphicx}
\usepackage{float}
\usepackage{bm}
\usepackage{breqn}
\usepackage{ctable}

\newcommand{\otoprule}{\midrule[\heavyrulewidth]}
\newcommand{\bs}[1]{\ensuremath{\boldsymbol{#1}}}
\newcommand{\zero}{\boldsymbol{0}}

\renewcommand{\u}{\bs{u}}
\renewcommand{\P}{\mathcal{P}}

\newcommand{\CO}{C_{\mathrm{O}}}
\newcommand{\CI}{C_{\mathrm{I}}}

\newcommand{\DO}{D_{\mathrm{O}}}
\newcommand{\DI}{D_{\mathrm{I}}}

\newcommand{\interleaver}{\pi}
\newcommand{\punct}{\mathcal{P}}

\newcommand{\MUX}{{\small \textsf{MUX}}}

\newcommand{\p}{\mathbf{p}}

\newcommand{\CPCCa}{\mathcal{C}_{\mathrm{PCC}}}
 
\newcommand{\CPCCSa}{\mathcal{C}_{\mathrm{SC-PCC}}}

\newcommand{\RSCCS}{R_{\mathrm{SC}}}
\newcommand{\KSCCS}{K_{\mathrm{SC}}}
\newcommand{\CSCCa}{\mathcal{C}_{\mathrm{SCC}}}
 
\newcommand{\CSCCSa}{\mathcal{C}_{\mathrm{SC-SCC}}}

\newcommand{\fs}[1]{\ensuremath{f_{\mathrm{#1},\mathrm{s}}}}
\newcommand{\fp}[1]{\ensuremath{f_{\mathrm{#1},\mathrm{p}}}}

\newcommand{\xsa}[2]{\ensuremath{p_{\mathrm{#1},\mathrm{#2}}}}
\newcommand{\ysa}[2]{\ensuremath{p_{\mathrm{#1},\mathrm{#2}}}}

\newcommand{\xs}[4]{\ensuremath{p_{\mathrm{#1},\mathrm{#2}}^{(#3,#4)}}}

\IEEEoverridecommandlockouts

\begin{document}

\title{Spatially Coupled Turbo Codes: Principles and Finite Length Performance}


\author{
\IEEEauthorblockN{Alexandre Graell i Amat$^\dag$, Saeedeh Moloudi$^\ddag$, and Michael Lentmaier$^\ddag$}
\IEEEauthorblockA{$\dag$Department of Signals and Systems, Chalmers University of Technology, Gothenburg, Sweden\\
  $\ddag$Department of Electrical and Information
  Technology, Lund University, Lund, Sweden\\
              alexandre.graell@chalmers.se,\{saeedeh.moloudi,michael.lentmaier\}@eit.lth.se}\\
              \thanks{This work was supported in part by the Swedish Research Council (VR) under grant \#621-2013-5477.}\vspace*{-1cm}
}


\maketitle

\begin{abstract}

In this paper, we give an overview of spatially coupled turbo codes (SC-TCs), the spatial coupling of parallel and serially concatenated convolutional codes, recently introduced by the authors. For presentation purposes, we focus on spatially coupled serially concatenated codes (SC-SCCs). We review the main principles of SC-TCs and discuss their exact density evolution (DE) analysis on the binary erasure channel. We also consider the construction of a family of rate-compatible SC-SCCs with simple $4$-state component encoders. For all considered code rates, threshold saturation of the belief propagation (BP) to the maximum a posteriori threshold of the uncoupled ensemble is demonstrated, and it is shown that the BP threshold approaches the Shannon limit as the coupling memory increases. Finally we give some simulation results for finite lengths.
\end{abstract}

\IEEEpeerreviewmaketitle

\section{Introduction}

Spatial coupling of low-density parity-check (LDPC) codes \cite{JimenezLDPCCC,LentmaierTransITOct2010,Kudekar_ThresholdSaturation} has revealed as a powerful technique to construct codes that universally achieve capacity for many channels under belief propagation (BP) decoding. The main principle behind this outstanding behavior is the convergence of the BP threshold to the maximum a posteriori (MAP) threshold of the underlying block code ensemble. This phenomenon, known as threshold saturation, was proven in \cite{Kudekar_ThresholdSaturation} for the binary erasure channel (BEC), and the proof was extended to binary-input memoryless symmetric channels in \cite{KuMeRiUr10}. More recently, threshold saturation has also been proven for nonbinary spatially coupled LDPC (SC-LDPC) codes \cite{AndGra13}. 

The concept of spatial coupling is not exclusive of LDPC codes, and also applies to other scenarios, such as relaying, compressed sensing, and statistical physics. In the realm of coding, spatial coupling has recently been applied to turbo-like codes \cite{MoloudiISIT14,MolLenGra14,Moloudi_SPCOM14}. In \cite{MoloudiISIT14}, an analysis of the BP threshold of braided convolutional codes (BCCs), a class of turbo-like codes introduced in \cite{ZhangBCC} which possess an inherent spatially coupled structure, was performed on the BEC. This structure has been later extended in \cite{Moloudi_SPCOM14} to larger coupling memories, and it was demonstrated that threshold saturation occurs. In \cite{MolLenGra14} the authors introduced the concept of spatially-coupled turbo codes (SC-TCs), as the spatial coupling of \textit{classical} turbo codes, i.e., Berrou \textit{et al.} and Benedetto \textit{et al.} parallel
and serially concatenated codes. It was shown that the BP threshold of SC-TCs improves with respect to that of the uncoupled ensembles and that it approaches the MAP threshold of the uncoupled turbo code as the coupling memory increases, suggesting threshold saturation.  

In this paper, we give an overview of SC-TCs. In particular, our main focus is on spatially-coupled serially concatenated codes (SC-SCCs). We first describe the SC-SCC ensemble introduced in \cite{MolLenGra14} and discuss the exact density evolution (DE) on the BEC. We then consider the construction of a family of rate-compatible SC-SCCs by applying puncturing. To this purpose, we consider the coupling of the class of serially concatenated codes (SCCs) initially introduced in \cite{GraMonVat05} and further analyzed in \cite{GraMonVat09,GraBraRas07ETT}. We give DE results for several code rates and compare the BP thresholds with that of spatially coupled parallel concatenated codes (SC-PCCs) and BCCs of the same rate. Threshold saturation to the MAP threshold is observed for all considered rates with large enough coupling memory. Furthermore, the DE results show performance very close to the Shannon limit. Finally, we give simulation results for finite lengths.

\section{Spatially Coupled Serially Concatenated Codes}
\label{sec:SC-SCCs}

\begin{figure*}[!t]
  \centering
    \includegraphics[width=\linewidth]{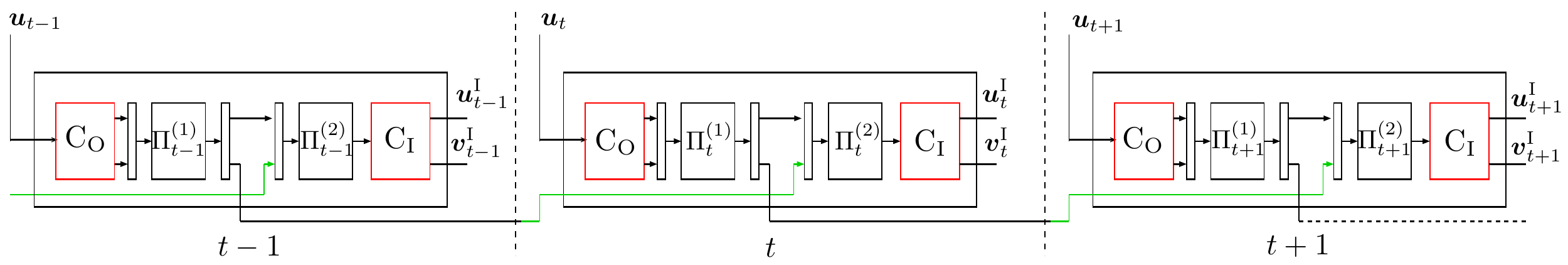}
\vspace{-4ex}
\caption{Block diagram of the encoder of a spatially coupled serially concatenated code with coupling memory  $m=1$.}
\label{fig:EncoderSCSCC}
\vspace{-2ex}
\end{figure*}

We consider the spatial coupling of rate $R=1/4$ SCCs, of length $K$ information bits, built from the concatenation of two identical rate$-1/2$ systematic recursive convolutional encoders, denoted by $\CO$ and $\CI$ (see Fig.~\ref{fig:EncoderSCSCC}). For simplicity, we describe spatial coupling with coupling memory $m=1$. An SC-SCC is built by applying a copy-and-coupling procedure, similar to that of SC-LDPC codes. Consider a chain of $L$ SCCs at time instants (spatial positions) $t\in[1,\ldots,L]$, as illustrated in Fig.~\ref{fig:EncoderSCSCC}, where $L$ is referred to as the coupling length. We denote by $\bs{u}_t$ the information sequence at time $t$. Also, denote by $\bs{v}_t^{\mathrm{O}}=(\bs{v}_t^{\mathrm{O,s}},\bs{v}_t^{\mathrm{O,p}})=(\bs{u}_t,\bs{v}_t^{\mathrm{O,p}})$ the code sequence at the output of the outer encoder $\CO$, where $\bs{v}_t^{\mathrm{O,s}}$ and $\bs{v}_t^{\mathrm{O,p}}$ correspond to the systematic and parity bits, respectively. Likewise, let $\bs{v}_t^{\mathrm{I,p}}$ be the encoded sequence at the output of the inner encoder $\CI$ corresponding to the parity bits. We also denote by $\tilde{\bs{v}}_t^{\mathrm{O}}$ the sequence $\bs{v}_t^{\mathrm{O}}$ rearranged by permutation $\Pi_t^{(1)}$.

An SC-SCC with $m=1$ is constructed by connecting each SCC in the chain to the one on the left and to the one on the right as follows. Divide randomly the sequence $\tilde{\bs{v}}_t^{\mathrm{O}}$ into two parts, $\tilde{\bs{v}}_{t,\text{A}}^{\mathrm{O}}$ and
$\tilde{\bs{v}}_{t,\text{B}}^{\mathrm{O}}$. Then, at time $t$ the
sequence at the input of the inner encoder $\CI$ is
$(\tilde{\bs{v}}_{t,\text{A}}^{\mathrm{O}},\tilde{\bs{v}}_{t-1,\text{B}}^{\mathrm{O}})$, properly reordered by permutation ${\Pi}_t^{(2)}$. 
In order to terminate the encoder of the SC-SCC, the information
sequences at the end of the chain are chosen in such a way that the
output sequence at time $t=L+1$ is
$\bs{v}_{L+1}^{\text{I}}=\bs{0}$. This entails a rate loss that vanishes as $L$ increases. A simple and practical way to terminate SC-SCCs is to set $\u_L=\zero$. This enforces $\bs{v}_{L+1}^{\text{I}}=\bs{0}$, since we can assume that $\u_t=\zero$ for $t>L$. Using this termination technique, only the parity bits of code $\CI$ need to be transmitted at time instant $L$. The code rate is therefore
\begin{equation}
\RSCCS=\frac{1}{4+\frac{2}{L-1}},
\end{equation}
where $\frac{2}{L-1}$ is the rate loss induced by the termination of the coupling chain. Note that $\lim_{L\rightarrow\infty}\RSCCS=R$. The information block length of the SC-SCC is $\KSCCS=(L-1)K$.

SC-PCCs can be constructed in a similar way. For details, the reader is referred to \cite{MolLenGra14}.

\subsection{Iterative decoding}

\begin{figure}[!t]
  \centering
    \includegraphics[width=.9\linewidth]{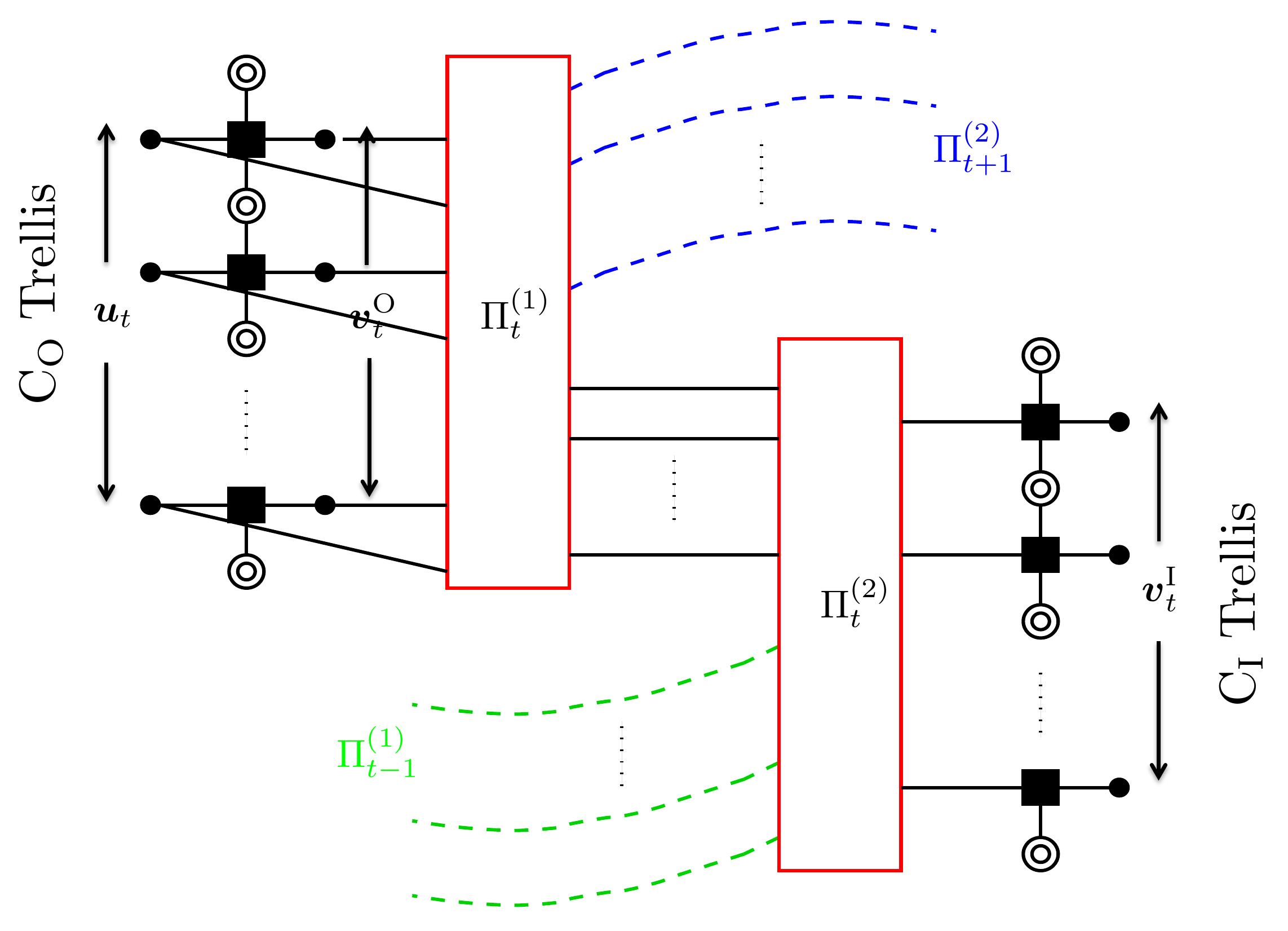}
\caption{Factor graph of a single section (time instant $t$) of a spatially coupled serially concatenated code.}
\label{factorS}
\vspace{-2ex}
\end{figure}

Similarly to turbo-like codes, SC-SCCs can be decoded using iterative message passing (belief propagation) decoding, where the component encoders of each SCC are decoded using the BCJR algorithm. The BP decoding of SC-SCCs can be easily visualized with the help of Fig.~\ref{factorS}, which depicts the factor graph of the decoder at time instant $t$ for $m=1$. We denote by $\DO$ and $\DI$ the decoder of the outer and inner encoder, respectively.

Decoder $\DO$ receives at its input information from the channel at time $t$ for both systematic and parity bits. Furthermore, it also receives a-priori information on both systematic and parity bits from
other decoders. As described above, at time $t$ the sequence $\CO$ is divided into two parts, $\tilde{\bs{v}}_{t,\text{A}}^{\mathrm{O}}$ and
$\tilde{\bs{v}}_{t,\text{B}}^{\mathrm{O}}$, which are encoded by the inner encoder in time instants $t$ and $t+1$, respectively. Correspondingly, $\DO$ receives a-priori information from $\DI$ at times $t$ (black lines connecting permutations $\Pi_t^{(1)}$ and $\Pi_t^{(2)}$) and $t+1$ (blue lines). Based on the information from the channel and from the companion decoders, $\DO$ computes the extrinsic information on the systematic and parity bits using the BCJR algorithm. Similarly, $\DI$ receives at its input information from the channel at time $t$ for the parity bits, and a-priori information on the input bits of $\CI$ from $\DO$ at times $t$ (black lines) and $t-1$ (green lines). 

\section{Density Evolution Analysis on the BEC}
\label{sec:DE}

We consider transmission over a BEC with erasure probability $\epsilon$, denoted by BEC$(\epsilon)$. In the following, we give the exact DE equations for $R=1/4$ SC-SCCs with $m=1$, derived in \cite{MolLenGra14}. We then discuss the modification of the equations to higher rates in Section~\ref{sec:RSCC}.

Let $\xsa{O}{s}$ and $\ysa{I}{s}$ be the average (extrinsic) erasure probability on the systematic bits at the output of the outer and inner decoder, respectively. Likewise, we define $\xsa{O}{p}$ and $\xsa{I}{p}$ for the parity bits at the output of the outer and inner decoder, respectively. The erasure probabilities $\ysa{I}{s}$ and $\ysa{I}{p}$ at time instant $t$ and decoding iteration $i$ can be written as
\begin{align}
\label{eq:LowerUpdates}
\xs{I}{s}{i}{t}&=\fs{I}\left(
q_{\text{O}}^{(i-1)},\epsilon\right)\\
\label{eq:LowerUpdatep}
\xs{I}{p}{i}{t}&=\fp{I}\left(
q_{\text{O}}^{(i-1)},\epsilon\right),
\end{align}
where
\begin{equation}
\label{eq:InnerUpdate3}
q_{\text{O}}^{(i-1)}=\epsilon \cdot\frac{\xs{O}{s}{i-1}{t}+\xs{O}{p}{i-1}{t}+\xs{O}{s}{i-1}{t-1}+\xs{O}{p}{i-1}{t-1}}{4},
\end{equation}
and $\fs{I}$ and $\fp{I}$ denote the inner decoder transfer functions for the systematic and parity bits, respectively. 

Likewise, $\ysa{O}{s}$ and $\ysa{O}{p}$ are given by
\begin{align}
\label{eq:OuterUpdates}
\xs{O}{s}{i}{t}&=\fs{O}\left(q_{\text{I}}^{(i-1)},q_{\text{I}}^{(i-1)}\right)\\
\label{eq:OuterUpdatep}
\xs{O}{p}{i}{t}&=\fp{O}\left(q_{\text{I}}^{(i-1)},q_{\text{I}}^{(i-1)}\right),
\end{align}
where
\begin{equation}
\label{eq:OuterUpdate3}
q_{\text{I}}^{(i-1)}=\epsilon \cdot \frac{\xs{I}{s}{i-1}{t}+\xs{I}{s}{i-1}{t+1}}{2} \ .
\end{equation}
and $\fs{O}$ and $\fp{O}$ denote the outer decoder transfer functions for the systematic and parity bits, respectively. 

The a-posteriori erasure probability on the information bits at time $t$ after
$i$ iterations is
\begin{equation}
\label{eq:appSCC}
p^{(i,t)}_{\rm a}=\epsilon \cdot \xs{O}{s}{i}{t} \cdot \frac{\xs{I}{s}{i}{t}+\xs{I}{s}{i}{t+1}}{2} \ .
\end{equation}

For the BEC, it is possible to derive analytical (exact) expressions for the transfer functions $\fs{O}$, $\fp{O}$, $\fs{I}$ and $\fp{I}$, using the method proposed in \cite{Kur03} and \cite{tenBrinkEXITConv} to compute the decoding erasure probability of convolutional encoders. DE is then performed by tracking the evolution of $p^{(i,t)}_{\rm a}$ in (\ref{eq:appSCC}) with the number of iterations, with the initialization $\xs{I}{s}{0}{t}=\xs{I}{p}{0}{t}=0$ for $t=0$ and $t>L$ and 1 otherwise, and
$\xs{O}{s}{0}{t}=\xs{O}{p}{0}{t}=0$ for $t=0$ and $t\ge L$ and 1 otherwise. The BP threshold corresponds to the maximum channel parameter $\epsilon$ such that $p^{(i,t)}_{\rm a}$ tends to zero for all time instants $t\in[1,\ldots,L]$ as $i$ tends to infinity, i.e., successful decoding is achieved.

Equations (\ref{eq:LowerUpdates})--(\ref{eq:appSCC}) can be easily generalized to larger coupling memories $m>1$. For details, we refer the reader to \cite{MolLenGra14}.

\section{Rate-Compatible Spatially Coupled Serially Concatenated Codes}
\label{sec:RSCC}

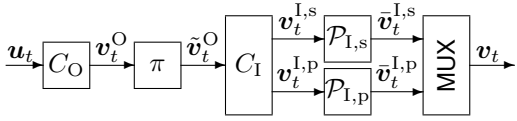
\begin{figure}[t]
\setlength{\unitlength}{1.0mm}%
  \begin{picture}(75,16)(1,13.5)
    \put(11.5,0){
    \put(0,20){\vector(1,0){5}} %
     \put(2,20.5){\makebox(0,0)[b]{$\u_t$}}
    \put(5,17){\framebox(6,6){$\CO$}}
    }
    \put(22.5,0){
    \put(0,20){\vector(1,0){6}} %
    \put(3,20.5){\makebox(0,0)[b]{$\bs{v}_t^{\mathrm{O}}$}}%
    \put(6,17){\framebox(6,6){$\interleaver$}}
    }
    \put(34.5,0){
    \put(0,20){\vector(1,0){6}} %
    \put(3,20.5){\makebox(0,0)[b]{$\tilde{\bs{v}}_t^{\mathrm{O}}$}}%
    \put(6,13.5){\framebox(6,13){$\CI$}}
    }
    \put(46.5,+3.5){
    \put(0,20){\vector(1,0){7}} %
    \put(3.5,20.5){\makebox(0,0)[b]{${\bs{v}}_t^{\mathrm{I,s}}$}}%
    \put(7,17){\framebox(6,6){${\P}_{\mathrm{I,s}}$}}
    \put(13,20){\vector(1,0){7}} %
    \put(16.5,20.5){\makebox(0,0)[b]{$\bar{\bs{v}}_t^{\mathrm{I,s}}$}}%
    }
    \put(46.5,-3.5){
    \put(0,20){\vector(1,0){7}} %
    \put(3.5,20.5){\makebox(0,0)[b]{${\bs{v}}_t^{\mathrm{I,p}}$}}%
    \put(7,17){\framebox(6,6){${\P}_{\mathrm{I,p}}$}}
    \put(13,20){\vector(1,0){7}} %
    \put(16.5,20.5){\makebox(0,0)[b]{$\bar{\bs{v}}_t^{\mathrm{I,p}}$}}%
    }
    \put(60.5,0){
    \put(6,13.5){\framebox(6,13){\rotatebox{90}{\MUX}}}
    \put(12,20){\vector(1,0){6}} %
    \put(14.5,20.5){\makebox(0,0)[b]{$\bs{v}_t$}}%
    }
  \end{picture}
\caption{Block diagram of the serially concatenated code.}
\label{fig:SystemModel_1}%
\vspace{-3ex}
\end{figure}

In this section, we construct a family of rate-compatible SC-SCCs by means of puncturing. In particular, we consider the coupling of the SCCs proposed in \cite{GraMonVat09,GraBraRas07ETT}.\footnote{This code structure was later generalized in \cite{GraRasBra11} to a code structure that encompasses both parallel concatenated codes and SCCs.} The block diagram of the SCC in \cite{GraMonVat09,GraBraRas07ETT} is depicted in Fig.~\ref{fig:SystemModel_1}, where we used the notation introduced in Section~\ref{sec:SC-SCCs} for the information and code sequences. In contrast to standard SCCs, characterized by the concatenation of an outer encoder with a rate $R_I\le 1$ inner encoder, where the outer encoder may be punctured to achieve higher rates, the SCCs proposed in \cite{GraMonVat09,GraBraRas07ETT} achieve high rates by moving the puncturing of the outer encoder to the inner encoder, which is punctured beyond the unitary rate. The result is that the interleaver gain is preserved for high rates, which results in codes that significantly outperform standard SCCs, especially for high rates.

In Fig.~\ref{fig:SystemModel_1}, we denote by ${\P}_{\mathrm{I,s}}$ and ${\P}_{\mathrm{I,p}}$ the puncturers for the systematic and parity bits, respectively, of $\CI$. A puncturer $\P$, applied to a code sequence, is commonly defined by a puncturing pattern $\p$ with pattern length $N_p$. For example, if $N_p = 4$ and puncturer $\P$ is chosen to puncture every fourth bit, the pattern is described as $\p = [1~1~1~0]$, where $0$ represents a punctured position. The puncturing pattern is then repeated periodically, and therefore $N_p$ is sometimes referred to as the puncturing period. A puncturer is also defined by a permeability rate $\rho\in[0,1]$, which gives the fraction of bits that survive after puncturing. For instance, for the example above $\rho=0.75$, i.e., $75\%$ of the bits survive after puncturing and the remaining $25\%$ are punctured. We use the notation $\bar{\bs{v}}$ to denote the punctured version of the codeword $\bs{v}$.
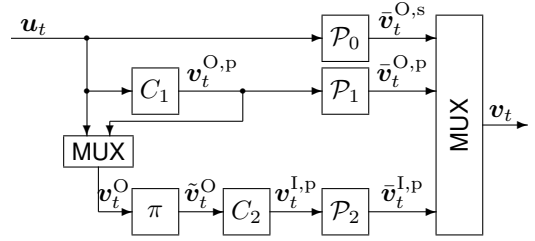
\begin{figure}[t]
\setlength{\unitlength}{1.0mm}%
  \begin{picture}(75,29)(-4,14)
    \put(18,0){
    \put(-12,40){\vector(1,0){41}} %
    \put(-9,40.5){\makebox(0,0)[b]{$\u_t$}}
    \put(29,37){\framebox(6,6){$\punct_0$}}
    \put(35,40){\vector(1,0){9}} %
    \put(39.5,40.5){\makebox(0,0)[b]{$\bar{\bs{v}}_t^{\mathrm{O,s}}$}}%
    \put(-2,40){\circle*{0.6}} %
    \put(-2,40){\vector(0,-1){13}} %
    }
    \put(18,-7){
    \put(-2,40){\circle*{0.6}} %
    \put(-2,40){\vector(1,0){6}} %
    \put(1,40.5){\makebox(0,0)[b]{}}%
    \put(4,37){\framebox(6,6){$C_1$}}
    \put(10,40){\vector(1,0){19}}%
    \put(14.5,40.5){\makebox(0,0)[b]{${\bs{v}}_t^{\mathrm{O,p}}$}}%
    \put(29,37){\framebox(6,6){$\punct_1$}}
    \put(35,40){\vector(1,0){9}} %
    \put(39.5,40.5){\makebox(0,0)[b]{$\bar{\bs{v}}_t^{\mathrm{O,p}}$}}%
    \put(18.5,40){\circle*{0.6}} %
    \put(18.5,40){\line(0,-1){4}} %
    \put(18.5,36){\line(-1,0){17.5}} %
    \put(1,36){\vector(0,-1){2}} %
    }
    \put(18,23){
    \put(-5,0){\framebox(9,4){\MUX}}%
    \put(-0.5,0){\line(0,-1){6}} %
    }
    \put(18,-23){
    \put(-0.5,40){\vector(1,0){4.5}} %
    \put(1.5,40.5){\makebox(0,0)[b]{$\bs{v}_t^{\mathrm{O}}$}}
    \put(4,37){\framebox(6,6){$\interleaver$}}
    \put(10,40){\vector(1,0){6}}%
    \put(13,40.5){\makebox(0,0)[b]{$\tilde{\bs{v}}_t^{\mathrm{O}}$}}
    \put(16,37){\framebox(6,6){$C_2$}}
    \put(25.5,40.5){\makebox(0,0)[b]{${\bs{v}}_t^{\mathrm{I,p}}$}}%
    \put(22,40){\vector(1,0){7}} %
    \put(26,40.5){\makebox(0,0)[b]{}}%
    \put(29,37){\framebox(6,6){$\punct_2$}}
    \put(35,40){\vector(1,0){9}} %
    \put(39.5,40.5){\makebox(0,0)[b]{$\bar{\bs{v}}_t^{\mathrm{I,p}}$}}%
    }
    \put(62,0){
    \put(0,14){\framebox(6,29){\rotatebox{90}{\MUX}}}%
    }
    \put(68,1.5){
    \put(0,27){\vector(1,0){6}} %
    \put(2.5,27.5){\makebox(0,0)[b]{$\bs{v}_t$}}%
    }
  \end{picture}
\caption{Equivalent block diagram of the serially concatenated code in Fig.~\ref{fig:SystemModel_1}.}%
\label{fig:SystemModel_2}%
\vspace{-3ex}
\end{figure}

For DE analysis and code optimization purposes, it is more convenient to consider the equivalent block diagram of Fig.~\ref{fig:SystemModel_2}, which allows to decouple the effects on the performance of the puncturing of information bits, outer encoder parity bits, and inner encoder parity bits. In Fig.~\ref{fig:SystemModel_2},
$\mathcal{C}_1$ and $\mathcal{C}_2$ are rate-$1$ encoders corresponding to the parity parts of $\CO$ and $\CI$, respectively, from Fig.~\ref{fig:SystemModel_1}. We denote by ${\P}_0$ and ${\P}_1$ the puncturers for the information bits and encoder $C_1$ code bits (i.e., encoder $\CO$ parity bits), respectively, and by ${\P}_2$ the puncturer for encoder $C_2$ code bits (i.e., encoder $\CI$ parity bits). The corresponding puncturing patterns are $\p_0$, $\p_1$ and $\p_2$. Note that the combination of ${\p}_0$ and ${\p}_1$ (properly reordered by the permutation $\pi$) translates into the equivalent ${\p}_{\mathrm{I,s}}$, while  ${\p}_1={\p}_{\mathrm{I,p}}$, where ${\p}_{\mathrm{I,s}}$ and ${\p}_{\mathrm{I,p}}$ correspond to the puncturers of Fig.~\ref{fig:SystemModel_1}. 
\begin{table*}[!t]
\caption{Thresholds for punctured spatially coupled turbo codes}
\vspace{-3.5ex}
\begin{center}\begin{tabular}{ccccccccc}
\toprule
Ensemble& Rate & $\rho_2$ & $\epsilon_{\text{BP}}$ & $\epsilon_{\text{MAP}}$  &$\epsilon^1_{\mathrm{SC}}$ & $\epsilon^3_{\mathrm{SC}}$ & $\epsilon^5_{\mathrm{SC}}$ & $\delta_{\rm{SH}}$ \\
\otoprule
$\CPCCa$/$\CPCCSa$ & $1/3$ & 1.0 & 0.6428 & 0.6553 & 0.6553 & 0.6553 & 0.6553 & 0.0113\\[0.5mm]
$\CSCCa$/$\CSCCSa$ & $1/3$ & 1.0 & 0.5405 & 0.6654 & 0.6437 & 0.6650 & 0.6654 & 0.0012\\[0.5mm]
\midrule
$\CPCCa$/$\CPCCSa$ & $1/2$ & 0.5 &  0.4606 & 0.4689 & 0.4689 & 0.4689 & 0.4689 & 0.0311\\[0.5mm]
$\CSCCa$/$\CSCCSa$ & $1/2$ & 0.5 & 0.3594 & 0.4981 & 0.4708 & 0.4975 & 0.4981 & 0.0019\\[0.5mm]
\midrule
$\CPCCa$/$\CPCCSa$ & $2/3$ & 0.25 &  0.2732 & 0.2772 & 0.2772 & 0.2772 & 0.2772 & 0.0561\\[0.5mm]
$\CSCCa$/$\CSCCSa$ & $2/3$ & 0.25 & 0.2038 & 0.3316 & 0.3303 & 0.3305 & 0.3315 & 0.0018\\[0.5mm]
\midrule
$\CPCCa$/$\CPCCSa$ & $3/4$ & 0.166 & 0.1854 & 0.1876 & 0.1876 & 0.1876 &  0.1876 & 0.0624\\[0.5mm]
$\CSCCa$/$\CSCCSa$ & $3/4$ & 0.166 & 0.1337 & 0.2486 & 0.2155 & 0.2471 & 0.2486 & 0.0014\\[0.5mm]
\midrule
$\CPCCa$/$\CPCCSa$ & $4/5$ & 0.125 & 0.1376 & 0.1391 & 0.1391 & 0.1391 & 0.1391 & 0.0609\\[0.5mm]
$\CSCCa$/$\CSCCSa$ & $4/5$ & 0.125 & 0.0942 & 0.1990 & 0.1644 & 0.1968 & 0.1989 & 0.0011\\[0.5mm]
\midrule
$\CPCCa$/$\CPCCSa$ & $9/10$ & 0.055 & 0.0578 & 0.0582 & 0.0582 & 0.0582 & 0.0582 & 0.0418\\[0.5mm]
$\CSCCa$/$\CSCCSa$ & $9/10$ & 0.055 & 0.0269 & 0.0996 & 0.0624 & 0.0930 & 0.0988 & 0.0012\\[0.5mm]
\bottomrule
\end{tabular} \end{center}
\label{Tab:BPThresholdsSCC} 
\vspace{-3ex}
\end{table*}

We denote by $\rho_0$, $\rho_1$ and $\rho_2$ the permeability rate of ${\P}_0$, ${\P}_1$ and ${\P}_2$, respectively. For the DE analysis, we consider random puncturing. Furthermore, we assume that $\rho_0=1$, i.e., the overall code is systematic. The code rate of the punctured SC-SCC is
\begin{equation}
\RSCCS=\frac{1}{(1+\rho_1+2\rho_2)+\frac{2}{L-1}},
\end{equation}
where $R=\frac{1}{1+\rho_1+2\rho_2}$ is the rate of the (uncoupled) punctured SCC, and $\frac{2}{L-1}$ is the rate loss induced by the termination of the coupling chain, which is independent of the code rate.

Assume that a code sequence $\bs{v}$ is randomly punctured with permeability rate $\rho$, and then transmitted over a BEC$(\epsilon)$. For the BEC, puncturing is equivalent to transmitting $\bs{v}$ through a BEC$(\epsilon_{\rho})$ resulting from the concatenation of two BECs, BEC$(\epsilon)$ and BEC$(1-\rho)$, where $\epsilon_{\rho}=1-(1-\epsilon)\rho$. The DE equations derived in the previous section can be easily modified to account for puncturing. The DE for punctured SC-SCCs is obtained by substituting $\epsilon\leftarrow\epsilon_{\rho_2}$ in (\ref{eq:LowerUpdates}), (\ref{eq:LowerUpdatep}), and modifying (\ref{eq:InnerUpdate3}) to
\begin{align*}
&q_{\text{O}}^{(i-1)}=\\
& \frac{\epsilon\cdot\left(\xs{O}{s}{i-1}{t}+\xs{O}{s}{i-1}{t-1}\right)+\epsilon_{\rho_1}\cdot\left(\xs{O}{p}{i-1}{t}+\xs{O}{p}{i-1}{t-1}\right)}{4},
\end{align*}
and (\ref{eq:OuterUpdates}), (\ref{eq:OuterUpdatep}) to
\begin{align}
\label{eq:OuterUpdatesPunct}
\xs{O}{s}{i}{t}&=\fs{O}\left(q_{\text{I}}^{(i-1)},{\tilde{q}}_{\text{I}}^{(i-1)}\right)\\
\label{eq:OuterUpdatepPunct}
\xs{O}{p}{i}{t}&=\fp{O}\left(q_{\text{I}}^{(i-1)},{\tilde{q}}_{\text{I}}^{(i-1)}\right),
\end{align}
where $q_{\text{I}}^{(i-1)}$ is given in (\ref{eq:OuterUpdate3}) and
\begin{equation}
\label{eq:OuterUpdate3Punct}
\tilde{q}_{\text{I}}^{(i-1)}=\epsilon_{\rho_1} \cdot \frac{\xs{I}{s}{i-1}{t}+\xs{I}{s}{i-1}{t+1}}{2} \ .
\end{equation}

\subsection{Puncturing Optimization}

For a given code rate $R$, the puncturing rates $\rho_1$ and $\rho_2$ may be optimized. In this paper, we consider the optimization of $\rho_1$ and $\rho_2$ such that the MAP threshold of the (uncoupled) SCC is maximized.\footnote{Note that for a given $R$ the optimization simplifies to the optimization of a single parameter, say $\rho_2$, since $\rho_1$ and $\rho_2$ are related by $\rho_1=\frac{1}{R}-1-2\rho_2$.} Alternatively, one may optimize $\rho_1$ and $\rho_2$ such that the BP threshold of the SC-SCC is optimized for a given coupling memory $m$. Rate-compatibility can be guaranteed by choosing $\rho_1$ and $\rho_2$ to be decreasing functions of $R$.

\subsection{Fixed Random Puncturing Pattern}

For finite lengths, random puncturing may not be optimal, since it may harm the distance properties of the punctured code. We therefore consider regular puncturing patterns. A rate-compatible code family can be defined by a series of nested puncturing patterns. For example, the
series $[1~1~1~1]$, $[1~1~1~0]$, $[0~1~1~0]$, $[0~1~0~0]$ represent a
rate-compatible code family with permeability rates $1,\frac{3}{4},\frac{1}{2},\frac{1}{4}$. It is also important to define a puncturing order $\dot{\p}$, i.e., the order in which the code bits will be punctured. For the example above, the puncturing order is $\dot{\p} = [4~1~3~2]$, since bit position 4
is punctured first, followed by positions 1 and 3. A specific puncturing pattern $\p$ is therefore defined by the corresponding pair $\{\dot{\p},\rho\}$, e.g., $\p = [1~1~1~0]\leftrightarrow\{\dot{\p},\rho\} = \{[4~1~3~2],\frac{3}{4}\}$. Here, we optimize the puncturing orders $\dot{\p}_1$ and $\dot{\p}_2$  to optimize the encoders distance properties, as proposed in \cite{GraMonVat09,GraBraRas07ETT}. Then, for a given code rate, the number of bits to be punctured in each puncturing pattern is chosen according to the optimal permeability rates to maximize the MAP threshold of the uncoupled ensemble. 

\section{Density Evolution Results}


In this section, we give numerical results for some rate-compatible SC-SCCs using the DE analysis described in Section~\ref{sec:DE}. For the examples in this section and in Section~\ref{sec:FL} we use simple $4$-state component encoders with generator polynomials $(1,5/7)$ in octal notation. In Table~\ref{Tab:BPThresholdsSCC}, we give the BP threshold of SC-SCCs ($\CSCCSa$)
for several code rates and coupling memory $m=1,3$ and $5$, denoted by $\epsilon^1_{\mathrm{SC}}$, $\epsilon^3_{\mathrm{SC}}$ and $\epsilon^5_{\mathrm{SC}}$, respectively. All thresholds correspond to the case $L\rightarrow \infty$. We also report in the table the BP threshold ($\epsilon_{\text{BP}}$) and the MAP threshold ($\epsilon_{\text{MAP}}$) of the uncoupled ensembles ($\CSCCa$). The MAP threshold was computed applying the area theorem \cite{Measson2009}. We observe that SCCs show a significant gap between the BP threshold and the MAP threshold, a well-known phenomenon for SCCs. The BP threshold is improved when spatial coupling is applied. In particular, a significant
improvement is observed for coupling memory $m=1$. The threshold can be further improved by increasing $m$. For all rates, the BP threshold approaches the MAP threshold of the uncoupled ensemble with increasing values of $m$, showing that threshold saturation occurs. 

In Table~\ref{Tab:BPThresholdsSCC} we also give the BP thresholds of
punctured SC-PCCs ($\CPCCSa$) \cite{MolLenGra14} with the same $4$-state component
encoders, where both encoders are equally punctured, which corresponds
to the optimal puncturing. For SC-PCCs, $\rho_2$ corresponds therefore
to the permeability rate of the puncturers applied to the parity bits
of both the upper and lower encoders of the parallel
concatenation. Uncoupled parallel concatenated codes (PCCs) show a
superior BP threshold as compared to SCCs. However, the MAP threshold
is poorer than that of SCCs and this gap increases for higher
rates. It is also well known that PCCs have poorer distance spectrum
compared to SCCs. 
Accordingly, in the coupled case the BP threshold of SC-SCCs is larger than that of SC-PCCs,
except for $R=1/3$ (in this case the PCC is not punctured) with
$m=1$.
It is interesting to note that despite the use of very simple
component encoders, the BP threshold of the considered SC-SCCs is very
close to the Shannon limit as $m$ increases (the last column of the
table gives the gap to the Shannon limit for the coupled ensembles
with $m= 5$, $\delta_{\rm{SH}}$).

\section{Finite Length Performance}
\label{sec:FL}

In this section, we provide simulation results for SC-SCCs with finite code lengths. In the simulations, we consider the spatial coupling of SCCs of length $K=1024$ information bits, $L=100$, and $m=1$. The information block  length of the SC-SCCs is $\KSCCS=101376$ bits. For simplicity, to construct the SC-SCCs we use the same S-random permutation for $\Pi^{(1)}$ in each time instant, and $\Pi^{(2)}$ is removed (see Fig.~\ref{fig:EncoderSCSCC}). The sequence at the input of $\CI$ at time instant $t$ is then obtained by alternating bits from $\tilde{\bs{v}}_{t,\text{A}}^{\mathrm{O}}$ and $\tilde{\bs{v}}_{t-1,\text{B}}^{\mathrm{O}}$. However, we remark that this is not necessarily the best choice and a careful optimization should be performed. The SC-SCCs are terminated by setting $\u_L=\zero$, as explained in Section~\ref{sec:SC-SCCs}. 
\begin{figure}[!t]
\centering
\includegraphics[width=1\columnwidth]{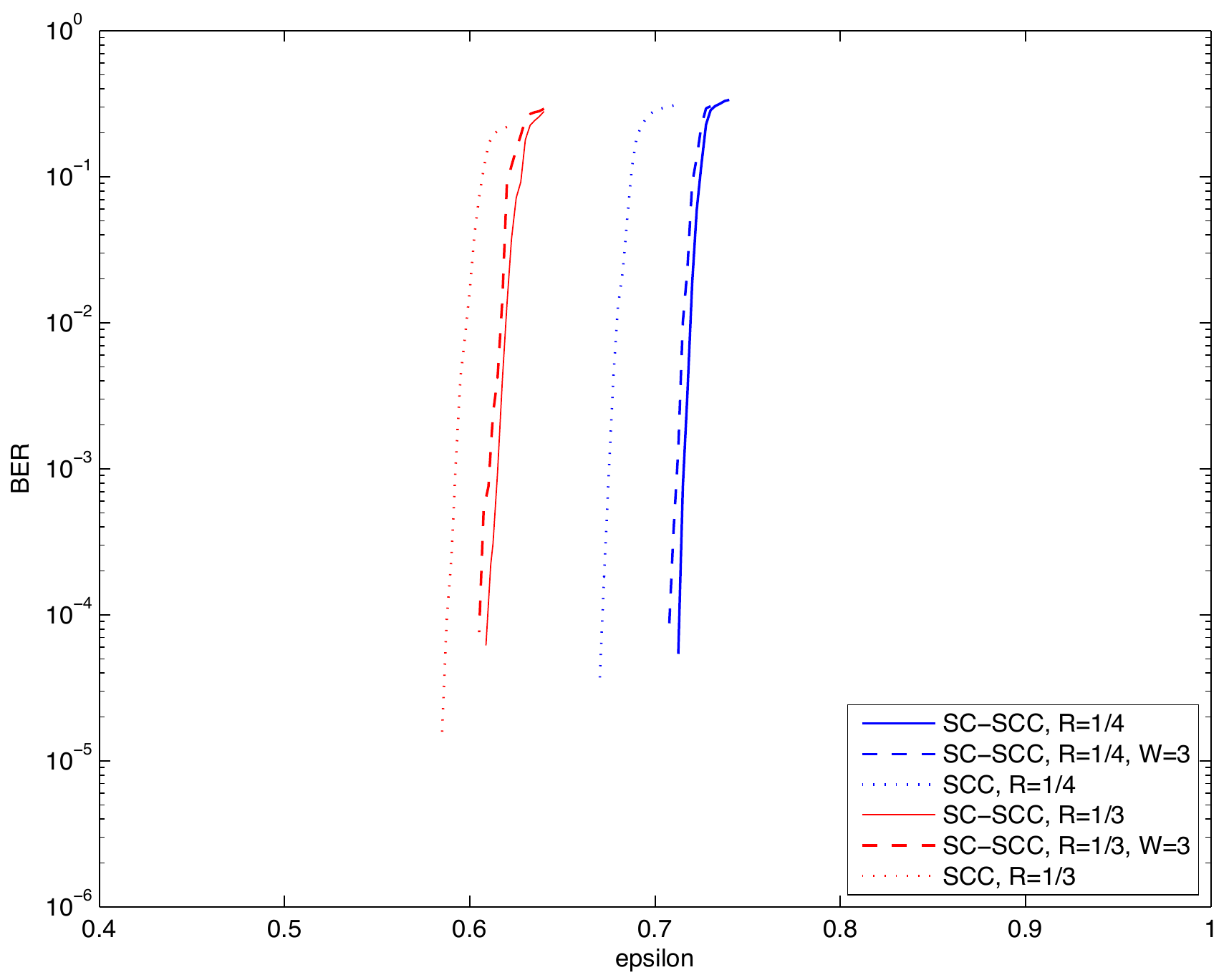}
\vspace{-0.75cm}
\caption{BER results for SC-SCCs with $L=100$ and $m=1$ on the binary erasure channel.}
\label{Fig:SimResults}
\end{figure}

In Fig.~\ref{Fig:SimResults} we give bit error rate (BER) results for SC-SCCs and rates $R=1/4$ (solid blue curve) and $R=1/3$ (solid red curve). For comparison purposes, we also plot the BER curve for the uncoupled ensembles (dotted curves) with $K=3072$.\footnote{For $R=1/3$, $\rho_2$ is set to 0.5, which maximizes the decoding threshold, but penalizes the error floor \cite{GraBraRas07ETT,GraRasBra11}. It is worth mentioning that the value that minimizes the error floor is $\rho_2=1$, which also maximizes the MAP threshold and is used for the coupled ensemble. Therefore, we expect the coupled ensemble to achieve a lower error floor.} In agreement with the DE results, a significant improvement is obtained by applying coupling. For comparison at low decoding latencies, we also plot in Fig.~\ref{Fig:SimResults} BER results using a sliding window decoder with window size $W=3$ (dashed curves). In this case, the decoding latency of the SC-SCCs is $K\cdot W=3072$ information bits, equal to the block length of the uncoupled ensembles. The SC-SCCs still significantly outperform their uncoupled counterparts.

\section{Conclusion}

In this paper, we gave an overview of spatially coupled turbo codes, the spatial coupling of parallel and serially concatenated codes, recently introduced by the authors. We also described the construction of a family of rate-compatible spatially coupled serially concatenated codes. Despite the use of simple 4-state component encoders, density evolution results demonstrate that performance very close to the Shannon limit can be achieved for all rates. Furthermore, threshold saturation of the BP threshold to the MAP threshold of the underlying uncoupled ensembles is observed for large enough coupling memory.

The invention of turbo codes and the rediscovery of LDPC codes, allowed to approach capacity with practical codes. Today, both turbo and LDPC codes are ubiquitous in communication standards. In the academic arena, however, the interest on turbo-like codes has been declining in the last years in favor of the more mathematically-appealing LDPC codes. The invention of spatially coupled LDPC codes and their capacity achieving performance have exacerbated this situation. Our SC-TCs and novel BCC ensembles demonstrate that other coding structures can also benefit from spatial coupling. Certainly, further research is required. Nevertheless, we believe that our works on spatially coupled turbo-like codes may contribute to regaining the interest in turbo-like coding structures.


\bibliographystyle{IEEEtran}

\end{document}